\documentclass[final,3p,times]{elsarticle}

\usepackage{amssymb}

\usepackage[cmex10]{amsmath}
\usepackage{amsfonts}

\usepackage{subfig}

\usepackage{bbm}
\usepackage[numbers]{natbib}
\usepackage{graphicx}
\usepackage{epstopdf}
\usepackage{url}

\usepackage{lineno}
\usepackage{color}
\usepackage{amsthm}

\usepackage{algorithm,algorithmic}

\journal{Communications in Nonlinear Science and Numerical Simulation}

\begin{document}

\begin{frontmatter}



\title{Optimal control of information epidemics modeled as Maki Thompson rumors}
\author{Kundan Kandhway\corref{cor1}}
\ead{kundan@dese.iisc.ernet.in}
\cortext[cor1]{Corresponding Author}
\author{Joy Kuri}
\ead{kuri@dese.iisc.ernet.in}
\address{Department of Electronic Systems Engineering, Indian Institute of Science, Bangalore 560012, India.}



\begin{abstract}

We model the spread of information in a homogeneously mixed population using the Maki Thompson rumor model. We formulate an optimal control problem, from the perspective of single campaigner, to maximize the spread of information when the campaign budget is fixed. Control signals, such as advertising in the mass media, attempt to convert ignorants and stiflers into spreaders. We show the existence of a solution to the optimal control problem when the campaigning incurs non-linear costs under the isoperimetric budget constraint. The solution employs Pontryagin's Minimum Principle and a modified version of forward backward sweep technique for numerical computation to accommodate the isoperimetric budget constraint. The techniques developed in this paper are general and can be applied to similar optimal control problems in other areas.

We have allowed the spreading rate of the information epidemic to vary over the campaign duration to model practical situations when the interest level of the population in the subject of the campaign changes with time. The shape of the optimal control signal is studied for different model parameters and spreading rate profiles. We have also studied the variation of the optimal campaigning costs with respect to various model parameters. Results indicate that, for some model parameters, significant improvements can be achieved by the optimal strategy compared to the static control strategy. The static strategy respects the same budget constraint as the optimal strategy and has a constant value throughout the campaign horizon. This work finds application in election and social awareness campaigns, product advertising, movie promotion and crowdfunding campaigns.

\end{abstract}

\begin{keyword}

Maki Thompson rumor model \sep Optimal control \sep Pontryagin's Minimum Principle \sep Social networks.


\end{keyword}

\end{frontmatter}



\section{Introduction}

Rumor models (\emph{e.g.} Daley Kendall, Maki Thompson) are used to model social contagion processes like spreading of information, ideas, fashion trends, etc. \cite{barrat2008dynamical, daley1964epidemics, maki1973mathematical} in a population. A piece of information affects human behavior which may be exploited by political and crowdfunding campaigners, companies for advertising their new products etc.. The goal of the campaigner is to reach as many people as possible by the campaign deadline, while making most efficient use of the available resources (\emph{e.g.} money, manpower). Since the contagion process is epidemic in nature, the allocation of resources over the campaign duration is important for optimal information spreading.

Rumor models are, in principle, similar to biological epidemic models like Susceptible-Infected-Susceptible (SIS) and Susceptible-Infected-Recovered (SIR), used for modeling the spread of pathogens in a population \cite{rapoport1953spread, goffman1964generalization}. The population is divided into three compartments (or classes): ignorants (those who don't have the information), spreaders (those who are spreading the information) and stiflers (those who have stopped spreading). Spreaders are generated at some rate due to ignorant-spreader contact---dynamics which is similar to the biological epidemic models. However the recovery process in the rumor models is different from that in biological epidemic models, which is explained in the following.

\emph{\textbf{Why rumor models are more appropriate than SIS/SIR models for information spreading:}} Recovery of a spreader to a stifler is spontaneous and independent of others in SIS/SIR models. To be specific: if the recovery rate is $\gamma$, then a spreader recovers after a time duration which is exponentially distributed with mean 1/$\gamma$, independent of her interaction with people in the population. On the other hand, in a rumor model, a spreader converts into a stifler at some rate, if she comes in contact with other spreaders or stiflers \cite{barrat2008dynamical, daley1964epidemics, maki1973mathematical}. For a recovery rate $\gamma$, the quantity 1/$\gamma$ provides a measure of the average number of interactions of a spreader with others, who are aware of the rumor, before she turns into a stifler \cite[Sec. 10.2]{barrat2008dynamical}. Information spreading is a psychological phenomenon, meeting others who already have the information changes the spreader's perception about information being new and she stops spreading. Due to this difference from SIS/SIR models, rumor models are more accurate in capturing information spreading dynamics. Similar arguments are applicable for fashion trends.

In this work, we aim to devise optimal information dissemination strategies, from the perspective of single campaigner, using the theory of optimal control. We assume that information spreading dynamics can be influenced by a control, which transfers individuals from the ignorant and stifler classes to the spreader class. Depending on the application---\emph{e.g.} political/crowdfunding campaigns; advertisement campaign for new products/services like smartphones, video games, satellite TV plans; fashion products like clothing, cosmetics---this can be done in various ways. Examples of ways in which control can be implemented in real systems include publishing manifestos, organizing political rallies/door-to-door campaigns, advertising in mass media and giving out discounts on new products, signing up brand ambassadors etc.. When an ignorant comes across the advertisement, she becomes aware of the information and starts spreading. Also, when a stifler sees the advertisement, her perception about an information being stale or fashion/product being old changes, and she starts spreading the information again or following the fashion trend again. Note that the control acts in addition to the epidemic ignorant-spreader contact which transfers ignorants to the spreader class.

Readers should not be misled by the term `rumor' in the Maki Thompson model. They can be used to model both useful and malicious information. In this work, we have used them for modeling only useful information and then attempt to maximize its reach. This paper does not address the problem of suppressing malicious information. Also note that apart from the direct applications listed in the previous paragraph, the tools and techniques developed in this paper can be used in other optimal control problems, such as mitigating the spread of biological epidemics and computer viruses, treating cancer \cite{fister1998optimizing} and suppressing corruption, terrorism and drug use \cite{grass2008optimal}.

\textbf{\emph{Related work and differences compared to the previous literature:}} Optimal control of SIS and SIR information epidemics was studied in \cite{karnik2012,kandhway2014run}. This work employs a more accurate information diffusion model, namely Maki Thompson model, as explained above. In addition, this work has an explicit budget constraint and non-linear resource application costs, which differentiates it from \cite{karnik2012,kandhway2014run}. The authors in \cite{belen2008,belen2005impulsive} studied impulsive control strategies to maximize Maki Thompson rumors. The rumor starts with a broadcast, and then there is an opportunity to trigger a second broadcast at a later stage. The work in \cite{belen2008,belen2005impulsive} determined the optimum time to trigger the second broadcast so that the number of ignorants in the system is minimized. In contrast, our formulation allows the system to be controlled throughout the campaign duration. In the applications considered---\emph{e.g.} political campaigns, product marketing, movie promotion---the campaigner tries to influence the system on a continuous basis. Advertisements appear in the mass media on a frequent basis, and not just once or twice; this motivates our model.

The work in \cite{sethi2008optimal} devises an optimal advertising and pricing plan for a newly launched product; however, it does not consider epidemic information diffusion in the population, as is the case in this paper. The authors in \cite{pittel1987spreading, chierichetti2009} analyze `push', `pull' and `push/pull' strategies for message diffusion, where nodes in the network either push the information to their neighbors, or pull it from them. Their aim is not to control the system but to find bounds on the number of communication rounds required to spread the information to almost all nodes in the network.

Optimal control of disease and computer virus epidemics is a well studied problem \cite{ledzewicz2011optimal,asano2008optimal,castilho2006optimal,behncke2000,gaff2009optimal,lashari2012optimal,morton1974,sethi1978,yan2008,zhu2012optimal,khouzani2011optimal}. In addition to the differences in the epidemic model, biological epidemics need to be contained, which is the opposite of spreading information. A biological epidemic has constant spreading rate, assuming the pathogens will not mutate within a season. In contrast, we have allowed the spreading rate to vary during the campaign duration to capture varying interest level of the population in the subject of the campaign during the time horizon of interest.

There is a sizeable literature on (uncontrolled) Maki Thompson rumor model and its extensions/generalizations. See for example \cite{belen2011classical,gani2000maki,pearce2000exact,lebensztayn2011behaviour,nekovee2007theory}.

\textbf{\emph{The following are the primary contributions of this paper:}}
\begin{enumerate}[(i)]
\item We have formulated and numerically solved the optimal control problem for maximizing information spread in the Maki Thompson model. In our formulation, the system can be controlled throughout the campaign duration, which is different from the impulsive control in \cite{belen2008,belen2005impulsive}. The control directly recruits ignorants and stiflers to spread the information. This can be done via methods such as placing advertisements in mass media. We assume a non-linear cost for applying the control and a fixed budget constraint. The standard forward backward sweep method used to solve the problem numerically needs to be modified due to the isoperimetric budget constraint.
\item Unlike previous literature, we have proved the existence of a solution to a system which has an isoperimetric budget constraint in the presence of non-linear costs. Standard Filippov/Cesari theorems are not applicable due to non-linear costs, because of which the control affects the system non-linearly, violating one of the requirements of the Filippov/Cesari theorem.
\item We have captured the varying interest level of the population in the subject of the campaign by a time-varying spreading rate over the campaign duration. In applications like poll campaigns, chatter about the election can be increase with time as the polling date approaches. In product marketing or movie promotion campaigns, the interest level of the population in the newly launched product/movie may decrease with time after its release. For applications such as crowdfunding or social awareness campaigns, the spreading rate is expected to be constant over time.
\end{enumerate}

The rest of this paper is organized as follows: Sec. \ref{sec:system_model_prob_formulation} formulates the optimal control problem. Sec. \ref{sec:existence} proves the existence of a solution to the optimal control problem formulated in Sec. \ref{sec:system_model_prob_formulation}. Sec. \ref{sec:solution} provides a framework to solve the problem formulated in Sec. \ref{sec:system_model_prob_formulation}. Sec. \ref{sec:results} discusses the results and Sec. \ref{sec:conclusion} concludes the paper.

\section{System model and problem formulation}
\label{sec:system_model_prob_formulation}

In this section we will briefly state the uncontrolled Maki Thompson model and use it to formulate an optimal control problem for the controlled system. Later in the section, we illustrate the developed model with a real world example of political campaigning. We have collected the definitions of all the parameters used in this paper in Table \ref{table:definition_parameters}.

\begin{table}[ht]
\caption{Definitions of parameters used in this paper}
\centering
\begin{tabular}{l l}
\hline
Symbol & Definition \\ [0.5ex]
\hline
$i(t)$ & fraction of ignorants in the population at time $t$ \\
$s(t)$ & fraction of spreaders in the population at time $t$ \\
$r(t)$ & fraction of stiflers in the population at time $t$ \\
$\beta_1(t)$ & per contact message spreading rate at time $t$ \\
$\gamma_1$ & per contact recovery rate \\
$k$ & number of other individuals an individual is in contact at any given time \\
$\beta(t) = k\beta_1(t)$ & spreading rate at time $t$ \\
$\gamma = k\gamma_1$ & recovery rate \\
$T$ & campaign deadline \\
$u(t)$ & control at time $t$ (\emph{e.g.} rate at which advertisements are put across in mass media) \\
$u_{max}$ & maximum allowed control, $0\leq u(t) \leq u_{max}$ \\
$c(u(t))$ & instantaneous cost incurred due to application of control \\
$B$ & budget \\
$b(t)$ & (cumulative) resource spent during $[0~~t]$ \\ [1ex] 
\hline
\end{tabular}
\label{table:definition_parameters}
\end{table}

Our objective is to minimize the number of ignorants at the end of the campaign. This is natural because we want to maximize the number of individuals who are aware of the information we are trying to spread.

\emph{\textbf{Uncontrolled Maki Thompson model:}} We consider a system with fixed population size. At time $t$, the fractions of ignorants, spreaders and stiflers in the population are represented by $i(t),~s(t)$ and $r(t)$ respectively, with $i(t)+s(t)+r(t)=1$. `Per contact message spreading rate' at time $t$ is $\beta_1(t)$ and `per contact recovery rate' is $\gamma_1$. They are interpreted in the subsequent paragraphs.

We first derive the rate of decrease of fraction of ignorants in the population at time $t$. In the beginning, at $t=0$, the system starts with $i(0)=1-s_0, s(0)=s_0, r(0)=0$, where $s_0$ is the initial fraction of spreaders which acts as the seed for the epidemic. `Per contact message spreading rate' can be interpreted as follows: in a small time interval $dt$ at time $t$, the message passes from a spreader to an ignorant due to a single ignorant-spreader contact with a probability $\beta_1(t)dt$. We assume that every member of the population interacts with an average of $k$ others at any time. Thus, an ignorant is in contact with (an average of) $ks(t)$ spreaders at time $t$. The message will be passed to the ignorant with probability, $1-(1-\beta_1(t)dt)^{ks(t)}\approx \beta_1(t)ks(t)dt$. Since the fraction of ignorants at time $t$ is $i(t)$ so the decrease in fraction of ignorants in small interval $dt$ at time $t$ is $\beta_1(t)ks(t)i(t)dt$. Defining $\beta(t)\triangleq \beta_1(t)k$, we get Eq. (\ref{eq:state_uncon_i}). In the rest of this paper, we refer to $\beta(t)$ as the `spreading rate'.

Now we derive the rate of increase of fraction of stiflers at time $t$. A spreader recovers to become a stifler due to interaction with other spreaders and stiflers. `Per contact recovery rate' is interpreted as follows: at any time $t$, any spreader in contact with a single spreader or stifler will convert to a stifler with probability $\gamma_1 dt$. Any member of the population interacts with $k$ others at any time. Hence, a spreader is in contact with an average of $k(s(t)+r(t))$ spreaders and stiflers, increasing her probability of recovery to $1-(1-\gamma_1 dt)^{k(s(t)+r(t))} \approx k(s(t)+r(t)) \gamma_1 dt$, in a small interval $dt$ at time $t$. Since the fraction of spreaders at time $t$ is $s(t)$, so the increase in fraction of spreaders at time $t$ in a small interval $dt$ is given by $s(t)k(s(t)+r(t)) \gamma_1 dt$. Defining $\gamma\triangleq \gamma_1 k$, we get the rate of increase of stiflers in the population as $\gamma s(t)(s(t)+r(t))$ (Eq. (\ref{eq:state_uncon_r})). We refer to $\gamma$ as the `recovery rate' in the rest of this paper. Eq. (\ref{eq:state_uncon_s}) is a consequence of Eqs. (\ref{eq:state_uncon_i}) and (\ref{eq:state_uncon_r}).

Thus, the evolution of the ignorants, spreaders and stiflers in the population in the uncontrolled Maki Thompson system is given by \cite[Sec. 10.2, adapted for time varying $\beta(t)$]{barrat2008dynamical}:
\begin{subequations}
\label{eq:uncontrolled_sys}
\begin{align}
\dot{i}(t) &= -\beta(t) i(t) s(t), \label{eq:state_uncon_i}\\
\dot{s}(t) &= \beta(t) i(t) s(t) - \gamma s(t)\big( s(t)+r(t) \big), \label{eq:state_uncon_s}\\
\dot{r}(t) &= \gamma s(t)\big( s(t)+r(t) \big). \label{eq:state_uncon_r}
\end{align}
\end{subequations}
These equations are mean field equations and are most accurate in the limit of large population size.

\emph{\textbf{Admissible controls:}} We denote the campaign deadline by $T$ and the set of all admissible controls by $U$. To define $U$, we first define the set, $\Psi$ of all equicontinuous functions over the campaign horizon $[0~~T]$. For all $\sigma \in \Psi$, $|\sigma(t)-\sigma(\hat t)| \leq C_\Psi(\epsilon)$, for $t,\hat t \in [0,T]$; $|t - \hat t| \leq \epsilon$, with $C_\Psi(\epsilon)\rightarrow 0$ as $\epsilon \rightarrow 0$ \cite[Sec. 1.6]{atkinson2009theoretical}. Then, any control signal 
\begin{align}
u\in U \triangleq \{\sigma\in \Psi:0\leq \sigma(t)\leq u_{max},~\forall t\in[0,T] \} \label{eq:def_U}
\end{align}
is admissible. Here $u_{max}$ is maximum allowed control strength. Practical considerations will require control signals to be bounded at all times. For example, there is a limit to how many advertisements one can place in newspapers in a day. We assume equicontinuity of the set $U$ as it aids in proving the existence of a solution to the optimal control problem. Note that assuming equicontinuity is milder than, for example, assuming differentiability and a large class of functions satisfy it.

\emph{\textbf{The controlled system:}} Notice that $\dot{i}(t)+\dot{s}(t)+\dot{r}(t)=0$, so only Eqs. (\ref{eq:state_uncon_i}) and (\ref{eq:state_uncon_s}) are sufficient to capture system dynamics, with $r(t) = 1-i(t)-s(t)$. System (\ref{eq:uncontrolled_sys}) can be controlled by a function $u\in U$ which transfers individuals from ignorant and stifler class to the spreader class as explained earlier. This is captured in Eqs. (\ref{eq:state_con_i}) and (\ref{eq:state_con_s}). We assume that application of the control incurs a non-linear cost, given by $c(u(t)) $ at time $t$. Also, we have a fixed budget, which is captured by Eq. (\ref{eq:budget}). The function $c(.)$ is assumed to be continuous and increasing in its argument. We want to maximize the number of individuals who are aware of the information by the campaign deadline $t=T$. We do not worry about the system evolution during $t<T$, so the reward function is $s(T)+r(T)=1-i(T)$. Hence we choose the cost function (to be minimized) to be $J=i(T)$ (Eq. (\ref{eq:cost_funtional})). The optimal control problem is:
\begin{subequations}
\label{eq:opt_control_prob}
\begin{align}
\underset{u\in U}{\textnormal{  min  }} J & = i(T), \label{eq:cost_funtional}\\
\text{subject to:~~} \dot i(t) & =  -\beta(t)i(t)s(t) - u(t)i(t), \label{eq:state_con_i} \\
\dot s(t) & = \big( \beta(t)+\gamma \big) i(t)s(t) - \gamma s(t) + u(t)i(t) + \alpha u(t) \big( 1-i(t)-s(t) \big), \label{eq:state_con_s} \\
s(0) & =  s_0,~~i(0)=1-s_0, \label{eq:init_cond} \\
&  \int_0^T c(u(t)) dt = B. \label{eq:budget}
\end{align}
\end{subequations}
Notice that we have removed the redundant equation involving $\dot r(t)$ and have made use of $r(t) = 1-i(t)-s(t)$ in Eq. (\ref{eq:state_con_s}). The factor $\alpha$ captures the effectiveness of the control on stiflers relative to ignorants. Also, Eq. (\ref{eq:budget}) can be replaced by an equivalent condition given by:
\begin{subequations}
\label{eq:budget_equivalent}
\begin{align}
\dot b(t) & = c(u(t)),\\
b(0) &=0,~~b(T)=B.
\end{align}
\end{subequations}
where $b(t)$ is a variable representing cumulative resource spent upto time $t$, $0\leq t\leq T$. This is a standard method to handle isoperimetric constraints in optimal control problems \cite[Sec. 16]{kamien1991dynamic}.

Notice that for $B\geq c(u_{max})T$, the optimum strategy is to apply the full strength control throughout the campaign duration. We will not discuss this case any further and will concentrate on more interesting---both practically and mathematically---resource constrained case, $B<c(u_{max})T$. Also notice that given a budget constraint, it is never optimal to underutilize the budget, hence we have used equality in Eq. (\ref{eq:budget}) without loss of generality.

\emph{\textbf{The example of political campaigning:}} In the strict sense, this example is discrete in nature, with system and control variables being discrete. But for modeling convenience, we use a continuous time model. This is common practice and all the examples cited in the `Related works' section do the same. These (mean field) models are accurate in the limit of large population size.

In this example, a political party wants to make the voters in a constituency aware of its candidate and manifesto. As noted before, in this work we are not studying the problem of suppressing misinformation/negative information about this party or candidate. The ignorants, whose fraction is denoted by $i(t)$ at time $t$, are not aware of the candidate. The spreaders, $s(t)$, are not only aware, but are also talking about this person whenever they get into conversation with others. And lastly, the stiflers, $r(t)$, have encountered enough number of spreaders and other stiflers---people who already have the information---to convince themselves that the information is no longer new, hence they have stopped spreading. They know about this candidate nonetheless. The polling day, which is the campaign deadline, is $T$ time units ahead. The spreading rate, $\beta(t)$ will be determined by the interest level of the population in talking about the upcoming election; a higher value means more interest. The reciprocal of the recovery rate, $1/\gamma$, is the measure of average number of spreaders/stiflers a spreader encounters before turning into a stifler.

The strategies employed by the political party to accelerate information diffusion act as the control here. These may be one, or a combination of the following strategies: advertising and publishing manifesto in newspapers/TV channels, door to door campaigns, SMS campaigns, distributing leaflets etc.. The intensity with which the above strategies are employed at time $t$ is denoted by $u(t)$, with $u_{max}$ denoting its maximum value which is enforced by physical limitations of the system.

The instantaneous monetary cost incurred due to the application of the above strategies is captured by the function $c(u(t))$ and the total money allocated by the party for this constituency is denoted by the budget $B$. The obvious goal here is to minimize the fraction of population who are not aware of the candidate at the polling day, $i(T)$. To do so, we need to select the best possible strategy, $u^*$ from the set of large number of available strategies, $U$. The solution to the optimal control problem provides $u^*$. If money is not a concern, the party would like to put across maximum possible numbers of advertisements, leaflets, SMSs etc. everyday before the deadline. But, as noted before, we are interested in the resource-constrained case where such a strategy is not feasible.

\section{Existence of a solution}
\label{sec:existence}
Proving the existence of a solution to the optimal control problem is of practical importance because not all problems admit their minimum/maximum, examples can be found in \cite[Chap. 3, Sec. 1]{fleming1975deterministic}. The standard methods to show the existence of a solution to an optimal control problem, Filippov/Cesari Theorems \cite[Chap. 3, Secs. 2 and 4]{fleming1975deterministic}, which were used in \cite{asano2008optimal,castilho2006optimal,behncke2000,gaff2009optimal} and many other works, are not applicable to Problem (\ref{eq:opt_control_prob}). This is because in Problem (\ref{eq:opt_control_prob})---with Eq. (\ref{eq:budget}) replaced by the equivalent condition (\ref{eq:budget_equivalent})---the control affects the system non-linearly. This is due to the non-linear function $c(.)$, which violates the linearity requirement of Filippov/Cesari Theorem. Hence we prove existence from first principles.

To this end, we make use of the fact that a continuous function defined on a compact set achieves its minimum/maximum \cite[Theorem 4.16]{rudin1964principles}, \emph{i.e.}, the solution exists. This is carried out in the following three steps:
\\ \emph{Step 1:} we define a compact set $W$ using the definition of $U$ in Eq. (\ref{eq:def_U}) and budget constraint in Eq. (\ref{eq:budget}).
\\ \emph{Step 2:} we show that $J$ in (\ref{eq:cost_funtional}) is continuous in the set $W$.
\\ \emph{Step 3:} we show that the solutions to the constraint Eqs. (\ref{eq:state_con_i}), (\ref{eq:state_con_s}) and (\ref{eq:init_cond}) exist for all elements in $W$.

\textbf{\emph{Step 1:}} Notice that the set $U$ being equicontinuous, equibounded and closed (all by definition) is compact \cite[Theorem 1.6.3]{atkinson2009theoretical}. Also, the set $V\triangleq \left\{\sigma:0\leq\sigma(t)\leq u_{max}, \int_0^Tc(\sigma(t)) dt =B\right\}$ is closed (because c(.) is continuous function by assumption). Since the intersection of compact and closed set is a compact set \cite[Pg. 38]{rudin1964principles}, hence, $W\triangleq U\cap V$ is compact.

The following problem is equivalent to Problem (\ref{eq:opt_control_prob}):
\begin{align}
\underset{u\in W}{\textnormal{  min  }} & J = i(T), \nonumber\\
\text{subject to:~~} & (\ref{eq:state_con_i}), (\ref{eq:state_con_s}) \textnormal{ and }(\ref{eq:init_cond}). \nonumber
\end{align}
It remains to show that $J$ is continuous in $W$ and constraints are satisfied for all $u\in W$. 

\textbf{\emph{Step 2:}} In fact the continuity of $J$ is valid at any $u\in U\supseteq W$. The function $J:U \rightarrow [0,1]$ is continuous at $u \in U$ if, for any $\epsilon>0$, we can find a $\delta>0$, such that $\big|J(u) - J(\hat u)\big| < \epsilon$, for all $\hat u \in U \cap \left\{ \hat p: |u - \hat p| < \delta \right\}$ \cite{rudin1964principles}. Let the system variables be denoted by $x(t)=\big(i(t),s(t)\big)$. The vector equation formed by combining Eqs. (\ref{eq:state_con_i}) and (\ref{eq:state_con_s}) is $\dot x(t)=X(x(t),t)$ where,
\begin{align}
X(x(t),t) = \begin{pmatrix}
\beta(t)i(t)s(t) - u(t)i(t) \\
\big( \beta(t)+\gamma \big) i(t)s(t) - \gamma s(t) + u(t)i(t) + \alpha u(t) \big( 1-i(t)-s(t) \big)
\end{pmatrix}. \nonumber
\end{align} The control signal $u$ is an explicit function of $t$, hence $X(.)$ which is a function of $x(t)$ and $u(t)$ is basically a function of $x(t)$ and $t$.

We use 1-norm for vectors and sup-norm for functions to measure the distance between them. We show the continuity of $\left.i(T)=i(t)\right|_{t=T}$ by invoking the `Theorem on Continuous Dependence' \cite[pg. 145]{walter1998ordinary} of the solution of an ordinary differential equation on the vector field $X(x(t),t)$. It states that if $x(t),~t\in[0,T]$, is a solution of $\dot x(t)=X(x(t),t)$; then given $\epsilon$ there exist $\delta$ such that $\big|x(t)-\hat x(t)\big| < \epsilon$, whenever $\big|X(x(t),t) - \hat X(x(t),t)\big| < \delta$, for $t\in[0,T]$. Here $\hat x(t)$ is the solution of perturbed version of $\dot x(t)=X(x(t),t)$, denoted by $\dot x(t)=\hat X(x(t),t)$, where $u$ in the vector field $X(x(t),t)$ is perturbed to $\hat u$ to get the perturbed vector field $\hat X(x(t),t)$.

We have,
\begin{align*}
|X(x(t),t)-\hat X(x(t),t)| &= \left|\big( u(t)-\hat u(t)\big)i(t) \right| + \left|\big( u(t)-\hat u(t)\big)i(t) + \alpha\big( u(t)-\hat u(t)\big)\big( 1-i(t)-s(t) \big) \right|\\
& \leq \left|\big( u(t)-\hat u(t)\big) \right| + \left|\big( u(t)-\hat u(t)\big)\right| + \left|\alpha\big( u(t)-\hat u(t)\big) \right| \\
& \leq (2+\alpha)|u - \hat u|.
\end{align*}
Note that $i(t)$ and $(1-i(t)-s(t))$ has maximum values 1. Thus,
\begin{align*}
& |u - \hat u|\leq \frac{\delta}{2+\alpha} \Rightarrow |X - \hat X| \leq \delta \Rightarrow |x(t) - \hat x(t)| \leq \epsilon, \forall t\in [0~T];
\end{align*}
which establishes the continuity of $i(t)|_{t=T}$ (a component of $x(t)$ evaluated at the final time $T$) in $u$.

In addition, the theorem on continuous dependence requires Lipschitz continuity of $X(.)$ in $x(t)$. Lipschitz continuity is also required for step 3 and is shown in the following.

\textbf{\emph{Step 3:}} The constrained initial value problem $\dot x(t)=X(x(t),t)$ (with initial conditions given by (\ref{eq:init_cond})) has a solution for any $u \in W$ if $X(x(t),t)$ is Lipschitz continuous in $x(t)$ for all $u \in W$ \cite[pg. 185]{birkhoff1989ordinary}. Again, Lipschitz continuity of $X(x(t),t)$ in $x(t)$ is valid for all $u\in U\supseteq W$. We denote $(\hat i(t), \hat s(t))$ by $\hat x(t)$ Notice that,
\begin{align*}
|X(x(t),t)-X(\hat x(t),t)|& = \left| \beta(t)\big( i(t)s(t)-\hat i(t)\hat s(t) \big) - u(t)\big( i(t)-\hat i(t) \big)\right| \\
& +  \left| \big(\beta(t)+\gamma\big)\big( i(t)s(t)-\hat i(t)\hat s(t) \big) - \gamma\big( s(t)-\hat s(t) \big) + u(t)\big( i(t)-\hat i(t) \big) - \alpha u(t)\big( i(t)-\hat i(t) + s(t)-\hat s(t) \big) \right| \\
& \leq \left| \beta(t)\big( i(t)s(t)-\hat i(t)\hat s(t) \big) \right| + \left|u(t)\big( i(t)-\hat i(t) \big)\right| +  \left| \big(\beta(t)+\gamma\big)\big( i(t)s(t)-\hat i(t)\hat s(t) \big) \right| \\
& + \left| \gamma\big( s(t)-\hat s(t) \big) \right| + \left| u(t)\big( i(t)-\hat i(t) \big) \right| + \left| \alpha u(t)\big( i(t)-\hat i(t) + s(t)-\hat s(t) \big) \right| \\
& \leq \bigg| \big(2\beta(t)+\gamma\big)\underbrace{\big( i(t)s(t)-\hat i(t)\hat s(t) \big)}_{i(t)s(t)-\hat i(t)s(t)+\hat i(t)s(t)-\hat i(t)\hat s(t)} \bigg| + \left|u(t)(2+\alpha)\big( i(t)-\hat i(t) \big)\right| + \left| \big(\gamma+\alpha u(t)\big)\big( s(t)-\hat s(t) \big) \right| \\
& \leq \left|\underset{t\in [0~~T]}{\textnormal{max}}\Big\{ 2\beta(t)+\gamma+u(t)(2+\alpha) \Big\} . \big( i(t)-\hat i(t) \big)\right| + \left| \underset{t\in [0~~T]}{\textnormal{max}}\big\{ 2\beta(t)+2\gamma+\alpha u(t)\big\} . \big( s(t)-\hat s(t) \big) \right| \\
& \leq \underset{t\in [0~~T]}{\textnormal{max}}\Big\{ 2\beta(t)+\gamma+u(t)(2+\alpha), 2\beta(t)+2\gamma+\alpha u(t) \Big\} . \left| \big( x(t)-\hat x(t) \big)\right|.
\end{align*}
Thus, $|X(x(t),t)-X(\hat x(t),t)|\leq C|x(t)-\hat x(t)|$ which establishes Lipschitz continuity of $X(x(t),t)$ in $x(t)$ for all $u \in U$.


\section{Solution to the optimal control problem}
\label{sec:solution}
In this section we first discuss the solution to Problem (\ref{eq:opt_control_prob})---with Eq. (\ref{eq:budget}) replaced by the equivalent condition (\ref{eq:budget_equivalent})---using Pontryagin's Minimum principle \cite{kamien1991dynamic}. This leads us to a system of ordinary differential equations (boundary value problem (BVP)) which are necessary conditions for optimum. Then we discuss a numerical method to solve the BVP. The standard forward backward sweep method used to solve the BVPs yielded by optimal control problems (see for example \cite{asano2008optimal, gaff2009optimal, lashari2012optimal}) is not directly applicable and needs to be adapted to take care of the isoperimetric budget constraint (\ref{eq:budget_equivalent}).

\subsection{Solution by Pontryagin's Minimum Principle}
We denote the adjoint variables by $\lambda_i(t),~\lambda_s(t)$ and $\lambda_b(t)$. At time $t$, $u^*(t)$ denotes the optimal control and, $i^*(t),s^*(t),b^*(t)$ and $\lambda_i^*(t), \lambda_s^*(t), \lambda_b^*(t)$ the state and adjoint variables evaluated at the optimum.
\\ \emph{Hamiltonian:} The Hamiltonian for Problem (\ref{eq:opt_control_prob}), with (\ref{eq:budget}) replaced by the equivalent condition (\ref{eq:budget_equivalent}), is given by,
\begin{align}
H(i(t),s(t),b(t),u(t),\lambda_i(t),\lambda_s(t),\lambda_b(t),t) &= \lambda_i(t)\left[\beta(t)i(t)s(t) - u(t)i(t)\right] \nonumber\\
&+ \lambda_s(t) \left[\big( \beta(t)+\gamma \big) i(t)s(t) - \gamma s(t) + u(t)i(t) + \alpha u(t) \big( 1-i(t)-s(t) \big)\right] \nonumber \\
&+ \lambda_b(t)\left[ c(u(t)) \right]. \nonumber
\end{align}
\emph{State equations:} Same as (\ref{eq:state_con_i}), (\ref{eq:state_con_s}), (\ref{eq:init_cond}) and (\ref{eq:budget_equivalent}) with $i(t),s(t),b(t),u(t)$ replaced by $i^*(t),s^*(t),b^*(t),u^*(t)$ respectively.
\\ \emph{Adjoint equations:} $\dot\lambda_i^*(t)$ is $- \frac{\partial}{\partial i(t)}H(i(t),s(t),b(t),u(t),\lambda_i(t),\lambda_s(t),\lambda_b(t),t)$ evaluated at the optimum.
\begin{align}
\dot\lambda_i^*(t)&= - \left.\frac{\partial}{\partial i(t)}H(i(t),s(t),b(t),u(t),\lambda_i(t),\lambda_s(t),\lambda_b(t),t) \right|_{\begin{smallmatrix}
i(t)=i^*(t), s(t)=s^*(t), b(t)=b^*(t), u(t)=u^*(t),\\
\lambda_i(t) = \lambda_i^*(t), \lambda_s(t) = \lambda_s^*(t), \lambda_b(t)=\lambda_b^*(t)
\end{smallmatrix}} \nonumber \\
&= \lambda_i^*(t)\beta(t)s^*(t) + \lambda_i^*(t)u^*(t) - \lambda_s^*(t)\beta(t)s^*(t) - \lambda_s^*(t)\gamma s^*(t) - \lambda_s^*(t)u^*(t) + \lambda_s^*(t)\alpha u^*(t). \label{eq:adjoint_lam_i}
\end{align}
Similarly,
\begin{align}
\dot\lambda_s^*(t)&= - \left.\frac{\partial}{\partial s(t)}H(i(t),s(t),b(t),u(t),\lambda_i(t),\lambda_s(t),\lambda_b(t),t) \right|_{\begin{smallmatrix}
i(t)=i^*(t), s(t)=s^*(t), b(t)=b^*(t), u(t)=u^*(t),\\
\lambda_i(t) = \lambda_i^*(t), \lambda_s(t) = \lambda_s^*(t), \lambda_b(t)=\lambda_b^*(t)
\end{smallmatrix}} \nonumber \\
&= \lambda_i^*(t)\beta(t)i^*(t) - \lambda_s^*(t)\beta(t)i^*(t) - \lambda_s^*(t)\gamma i^*(t) + \lambda_s^*(t)\gamma + \lambda_s^*(t)\alpha u^*(t). \label{eq:adjoint_lam_s}
\end{align}
And,
\begin{align}
\dot\lambda_b^*(t)&= - \left.\frac{\partial}{\partial b(t)}H(i(t),s(t),b(t),u(t),\lambda_i(t),\lambda_s(t),\lambda_b(t),t) \right|_{\begin{smallmatrix}
i(t)=i^*(t), s(t)=s^*(t), b(t)=b^*(t), u(t)=u^*(t),\\
\lambda_i(t) = \lambda_i^*(t), \lambda_s(t) = \lambda_s^*(t), \lambda_b(t)=\lambda_b^*(t)
\end{smallmatrix}} = 0. \label{eq:adjoint_lam_b}
\end{align}
\emph{Hamiltonian minimizing condition:} At the interior points,
\begin{align}
& \left.\frac{\partial}{\partial u(t)} H(i(t),s(t),b(t),u(t),\lambda_i(t),\lambda_s(t),\lambda_b(t),t)\right|_{\begin{smallmatrix}
i(t)=i^*(t), s(t)=s^*(t), b(t)=b^*(t), u(t)=u^*(t),\\
\lambda_i(t) = \lambda_i^*(t), \lambda_s(t) = \lambda_s^*(t), \lambda_b(t)=\lambda_b^*(t)
\end{smallmatrix}} \nonumber \\
& = -\lambda_i^*(t)i^*(t) + \lambda_s^*(t)i^*(t) + \lambda_s^*(t)\alpha\big( 1-i^*(t)-s^*(t) \big) + \lambda_b^*(t)c'(u^*(t)) = 0. \nonumber
\end{align}
Hence this condition leads to,
\begin{align}
u^*(t) &=
\begin{cases}
0 
& \text{if } c'^{-1}\left( \frac{\lambda_i^*(t)i^*(t)-\lambda_s^*(t)i^*(t)-\lambda_s^*(t)\alpha (1-i^*(t)-s^*(t))}{\lambda_b^*(t)} \right)<0, \\
c'^{-1}\left( \frac{\lambda_i^*(t)i^*(t)-\lambda_s^*(t)i^*(t)-\lambda_s^*(t)\alpha (1-i^*(t)-s^*(t))}{\lambda_b^*(t)} \right) 
& \text{if } 0\leq c'^{-1}\left( \frac{\lambda_i^*(t)i^*(t)-\lambda_s^*(t)i^*(t)-\lambda_s^*(t)\alpha (1-i^*(t)-s^*(t))}{\lambda_b^*(t)} \right) \leq u_{max}, \\
u_{max} 
& \text{if } c'^{-1}\left( \frac{\lambda_i^*(t)i^*(t)-\lambda_s^*(t)i^*(t)-\lambda_s^*(t)\alpha (1-i^*(t)-s^*(t))}{\lambda_b^*(t)} \right)>u_{max},
\end{cases} \nonumber \\
\Rightarrow u^*(t) & = \text{max} \left\{0, \text{min}\left\{u_{max}, c'^{-1}\left( \frac{\lambda_i^*(t)i^*(t)-\lambda_s^*(t)i^*(t)-\lambda_s^*(t)\alpha (1-i^*(t)-s^*(t))}{\lambda_b^*(t)} \right) \right\} \right\}. \label{eq:control_in_st_ad_var}
\end{align}
\emph{Transversality conditions:} The transversality conditions yield,
\begin{align}
\lambda_i^*(T) = 1, \lambda_s^*(T)=0, \lambda_b^*(T)=\text{free}. \label{eq:transversality_cond}
\end{align}

\subsection{Numerical solution and issues in computation}
To solve the optimal control problem numerically, we have to solve the BVP involving state and adjoint equations, also called the optimality system. The state equations are given by (\ref{eq:state_con_i}), (\ref{eq:state_con_s}), (\ref{eq:init_cond}) and (\ref{eq:budget_equivalent}); and the adjoint equations are given by (\ref{eq:adjoint_lam_i}), (\ref{eq:adjoint_lam_s}), (\ref{eq:adjoint_lam_b}) and (\ref{eq:transversality_cond}). Note that the value of the control variable (from Eq. (\ref{eq:control_in_st_ad_var})) has to be substituted in the above mentioned differential equations to get a system entirely in terms of state and adjoint variables.

The optimality system may be solved using boundary value problem solving techniques such as the shooting method \cite[Sec. 1.3, 1.4]{kutz2005practical}. But we found that the naive implementation of the shooting algorithm stalls before converging to a correct solution---possibly because of inaccuracies in the numerically computed gradient values. Also, due to the isoperimetric constraint, $b(0)=0$ and $b(T)=B$ in (\ref{eq:budget_equivalent}), it is not possible to implement naive forward backward sweep algorithm. Hence we briefly discuss the adaptation of the forward backward sweep algorithm which was used to solve the optimality system in the following.

Due to (\ref{eq:adjoint_lam_b}) and (\ref{eq:transversality_cond}), $\lambda_b^*(t)$ is an unknown value which is constant over time, $0\leq t\leq T$, for the optimality system, call it $\lambda_b^{c*}$. We have taken the approach of finding $\lambda_b^{c*}$ using bisection algorithm. Initialize the computation with two approximate values of $\lambda_b^{c*}$ (call them $\lambda_{b-high}^{c*}$ and $\lambda_{b-low}^{c*}$), one for which $b(T)<B$, and other for which $b(T)>B$. Then, refine the value of $\lambda_b^{c*}$ using bisection method till the constraint $b(T)=B$ is satisfied with desired tolerance. Details are in Algorithm \ref{alg:fw_back_sweep}.

\renewcommand{\algorithmiccomment}[1]{\hfill\emph{\% #1}}
\renewcommand{\algorithmicrequire}{\textbf{Input:}}
\renewcommand{\algorithmicensure}{\textbf{Output:}}
\begin{algorithm}
\caption{Modified forward backward sweep algorithm.}
\label{alg:fw_back_sweep}
\begin{algorithmic}[1]
	\REQUIRE $\lambda_{b-low}^{c*},~\lambda_{b-high}^{c*},~B_{th},~\lambda_{th},~B,~N_{sweep}$ (and other inputs $T$, $\beta(t)~\forall t\in[0,T]$, $\gamma$, $u_{max}$, $s_0$ and $\alpha$).
	\ENSURE The optimal control signal $u^*(t)$.
	\REPEAT
		\STATE $\lambda_b^{c*} \leftarrow (\lambda^{c*}_{b-low}+\lambda^{c*}_{b-high})/2$
		\STATE $u^*(t)\leftarrow 0,~\forall t\in[0,T]$
		\FOR{$j=1$\TO $N_{sweep}$}
			\STATE Calculate $i^*$ and $s^*$ using state equations (\ref{eq:state_con_i}) and (\ref{eq:state_con_s}) with initial conditions $i^*(0)=1-s_0$ and $s^*(0)=s_0$. \COMMENT{forward sweep}
			\STATE Calculate $\lambda_i^*$ and $\lambda_s^*$ using adjoint equations (\ref{eq:adjoint_lam_i}) and (\ref{eq:adjoint_lam_s}) with terminal conditions $\lambda_i^*(T)=1$ and $\lambda_s^*(T)=0$. \COMMENT{backward sweep}
			\STATE Calculate $u^*$ using (\ref{eq:control_in_st_ad_var}).
		\ENDFOR
		\STATE $b_{\lambda_b^{c*}}(T) \leftarrow \int_0^Tc(u^*(t))dt$
		\IF{$b_{\lambda_b^{c*}}(T)>B$}
			\STATE $\lambda_{b-low}^{c*} \leftarrow \lambda_b^{c*}$
		\ENDIF
		\IF{$b_{\lambda_b^{c*}}(T)<B$}
			\STATE $\lambda_{b-high}^{c*} \leftarrow \lambda_b^{c*}$
		\ENDIF
	\UNTIL{$\bigg(\big|b_{\lambda_b^{c*}}(T) - B\big|<B_{th}\bigg)$ and $\bigg(\big|\lambda_{b-high}^{c*}-\lambda_{b-low}^{c*}\big|<\lambda_{th}\bigg)$}.
\end{algorithmic}
\end{algorithm}

The values of $\lambda_{b-low}^{c*},~\lambda_{b-high}^{c*},~B_{th},~\lambda_{th}$ and $N_{sweep}$ used in all of the computations in this paper are 0, 100, $10^{-4}$, $10^{-4}$ and 50 respectively. Since $\lambda_{b-low}^{c*}$ is small, so control computed by (\ref{eq:control_in_st_ad_var}) is large, hence $b_{\lambda_{b-low}^{c*}}(T)$ is large (very close to maximum allowed budget, $c(u_{max})T$). Similarly $\lambda_{b-high}^{c*}$ is large, so control computed by (\ref{eq:control_in_st_ad_var}) is small, hence $b_{\lambda_{b-high}^{c*}}(T)$ is small (very close to 0). These values were found to be suitable to initialize the bisection method.

We have implemented Algorithm \ref{alg:fw_back_sweep} in MATLAB and have used its initial value problem solver \texttt{ode45()} to evaluate the differential equations. The solver uses fourth order Runge-Kutta algorithm with variable step size for computation and is capable of integrating backwards as required by the adjoint equations.

\section{Results}
\label{sec:results}
We first discuss the shapes of the control signal for constant and variable spreading rate profiles in Secs. \ref{sec:result_const_beta} and \ref{sec:reslut_variable_beta} respectively. Depending on the application, the spreading and recovery rate of the information epidemic may vary a lot. This depends on interest of people in conversing about the topic in question. Thus, we have used different parameter values in Secs. \ref{sec:result_const_beta} and \ref{sec:reslut_variable_beta} to model epidemics of varying virulence. The shape of the control signals varies considerably when the values of spreading and recovery rates are changed, as will be seen in Figs. \ref{fig:how_control_look_constant_beta} and \ref{fig:how_control_look_variable_beta}. Later in Sec. \ref{sec:result_effect_param}, we discuss the variation in the cost function (\ref{eq:cost_funtional}) with respect to various model parameters and compare the performance of the optimal control with the static control. In this paper we have assumed the cost of application of control to be, $c(u(t))=u^2(t)$.

\subsection{Shape of the control signal, constant spreading rate}
\label{sec:result_const_beta}

\begin{figure}[ht!]
\subfloat[Shapes of the control signals. \label{fig:how_control_look_constant_beta}]{
\includegraphics[width=80mm]{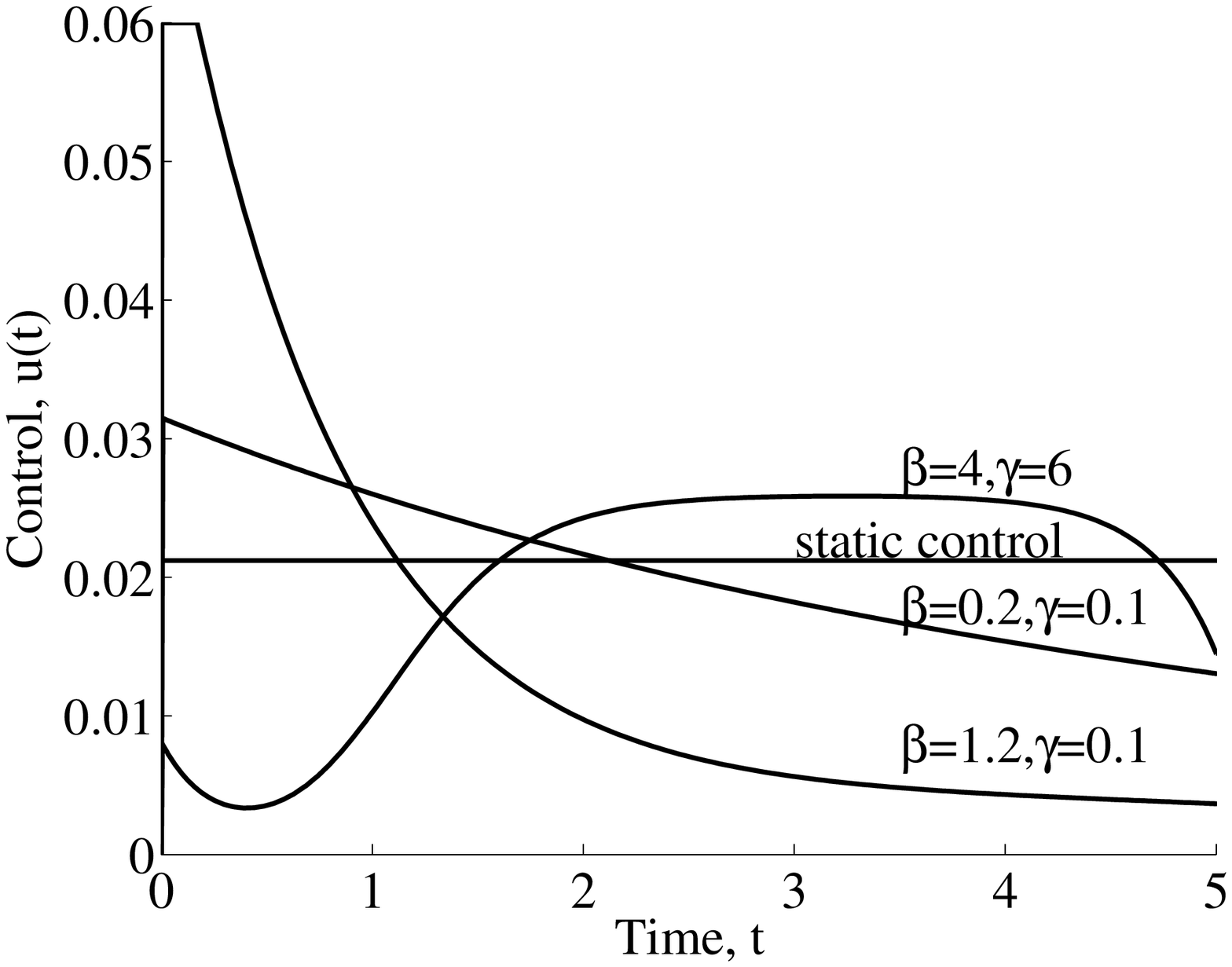} }
\hfill
\subfloat[Cumulative resource spent by the optimal and static controls. \label{fig:cumulative_resource_spent_constant_beta}]{
\includegraphics[width=80mm]{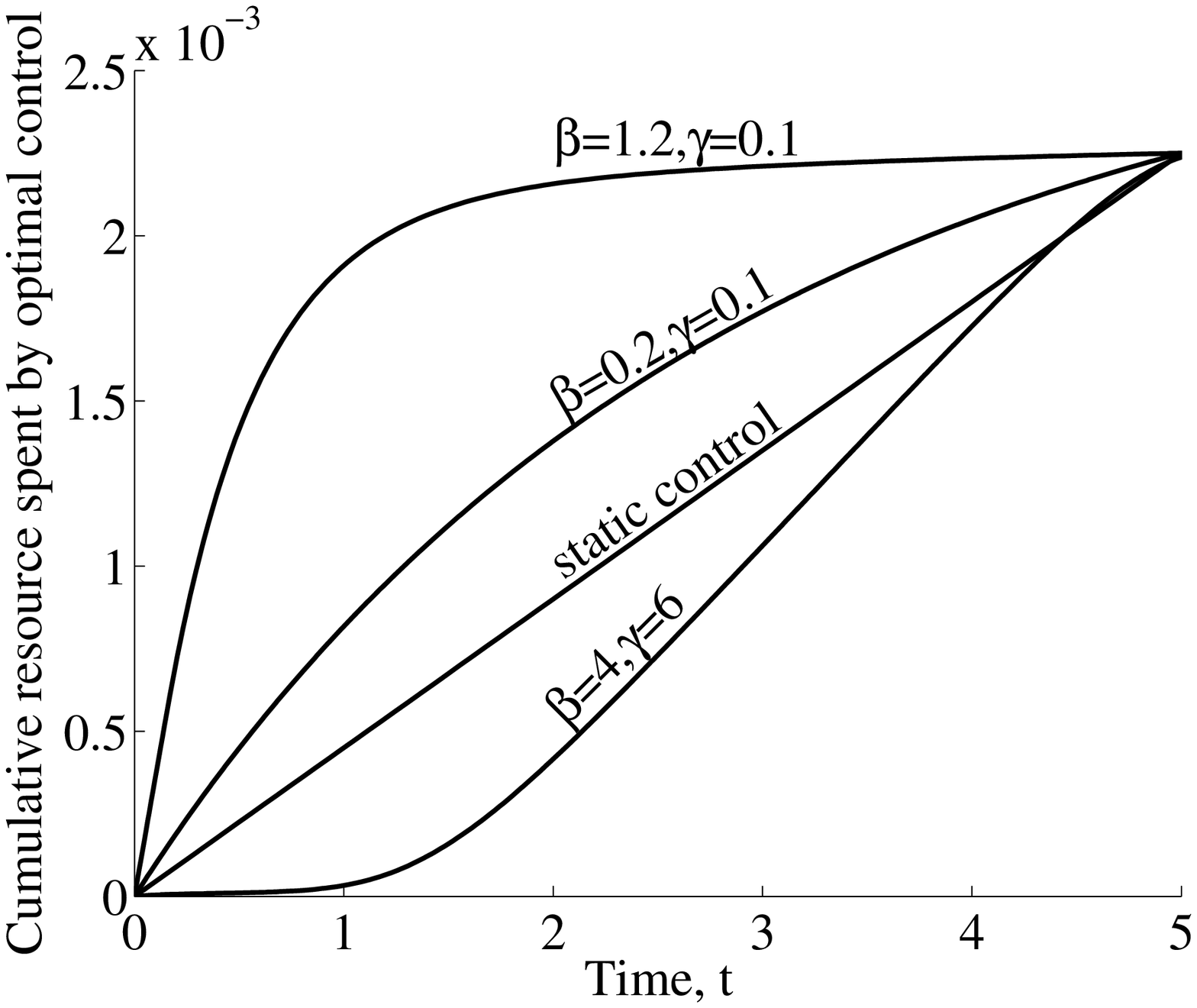} }
\caption{Parameter values: $c(u(t))=u^2(t),T=5,u_{max}=0.06,B=u_{max}^2T/8,s_0=0.01, \alpha = 0.5$.}
\label{fig:control_cumulative_resource_constant_beta}
\end{figure}

We first study the shape of the control signal when the spreading rate profile is constant over time, \emph{i.e.} $\beta(t)=\beta$, for $t\in[0~~T]$. The parameter values used to generate the given figure can be found in its caption. The campaign deadline is fixed at $T=5$, maximum control effort $u_{max}=0.06$, effectiveness of the control on stiflers $\alpha=0.5$, initial fraction of spreaders $s_0=0.01$ and budget $B=u_{max}^2T/8$ (resource constrained case). Since we have assumed $c(u(t))=u^2(t)$, and $0\leq u(t)\leq u_{max}$, this value of $B$ corresponds to one-eighth of the maximum possible value.

In contrast to SIS/SIR models, the Maki Thompson model does not have an epidemic threshold \cite[Sec. 10.2]{barrat2008dynamical}. Whenever $\beta/\gamma>0$, which is always satisfied for non zero $\beta$, the information epidemic kicks off. In Fig. \ref{fig:how_control_look_constant_beta} we have shown the shapes of the control signals for three different values of the spreading rates $\beta$ and recovery rates $\gamma$. Our formulation aims to minimize the fraction of ignorants at the end of the campaign deadline (Eq. (\ref{eq:cost_funtional})), hence the application of the optimal strategy leads to lower value of $i(T)$ when compared to the non-optimal strategies where no control or a static control is applied. The static control respects the same budget as the optimal strategy and has constant value throughout the campaign horizon.

For strong epidemic with slow recovery ($\beta=1.2,\gamma=0.1$) it is advisable to exert more control effort at the beginning stages of the epidemic. Doing so increases the spreaders at early stages of the campaign, slow recovery means the population is sustained in spreading state, which helps in further information dissemination. For this case, the final fraction of ignorants with optimal strategy, static strategy and no control are 0.0697, 0.0909 and 0.2150 respectively showing the effectiveness of the optimal strategy over the static strategy.

Fig. \ref{fig:cumulative_resource_spent_constant_beta} shows the cumulative budget spent over time by the optimal strategy, $b^*(t)=\int_0^tc(u^*(t))dt$. It gives an idea of how monetary expenditure should be planned over the time horizon of interest. To implement the optimal strategy, either follow the advice in Fig. \ref{fig:how_control_look_constant_beta} for the intensity with which the control (\emph{e.g.} advertisements in mass media) is applied to the system; or alternatively allocate $b^*(t_2)-b^*(t_1)$ amount of resource during time interval $[t_1~~t_2]$.

The variability in the strength of the control signal over time decreases for mild epidemics with slow spreading and recovery ($\beta=0.2,\gamma=0.1$). In this case a more uniform control is desired to keep the population of spreaders large enough throughout the campaign duration. If a large portion of the budget is used in the early stages like the previous case, the recovery of individuals will reduce the number of spreaders in middle and final stages of the campaign, which is counter-productive. Since spreading is slow, not enough new spreaders are generated. For this case, the final fraction of ignorants with optimal strategy, static strategy and no control are 0.6121, 0.8178 and 0.9733 respectively.

Shift in the control effort to later stages of campaign is even more prominent when recovery rate is too high for the given campaign deadline ($\beta=4,\gamma=6$). In this case, control strength is very small in the initial stages, however it picks up in the middle stages before reducing again at the final stages.

\subsection{Shape of the control signal, variable spreading rate}
\label{sec:reslut_variable_beta}

\begin{figure}[ht!]
\centering
\includegraphics[width=80mm]{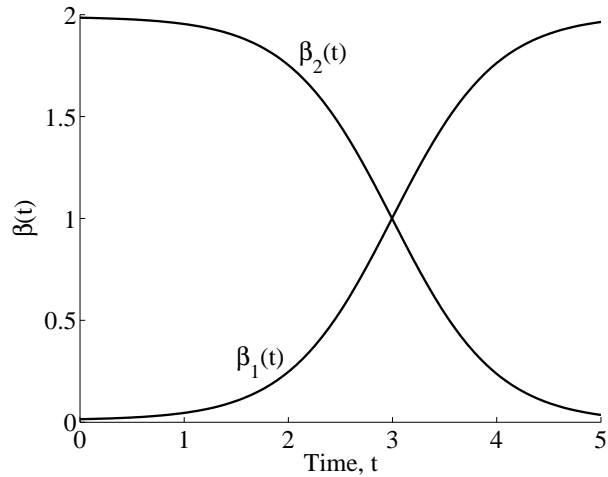}
\caption{Variable spreading rate profiles given in Eq. (\ref{eq:var_beta}). Parameter values: $\beta_m=.01, \beta_M=2, T=5, a_1=2, c_1=3, a_2=2, c_2=2$.}
\label{fig:variable_beta}
\end{figure}

\begin{figure}[ht!]
\subfloat[Shapes of the control signals for variable spreading rates. \label{fig:how_control_look_variable_beta}]{
\includegraphics[width=80mm]{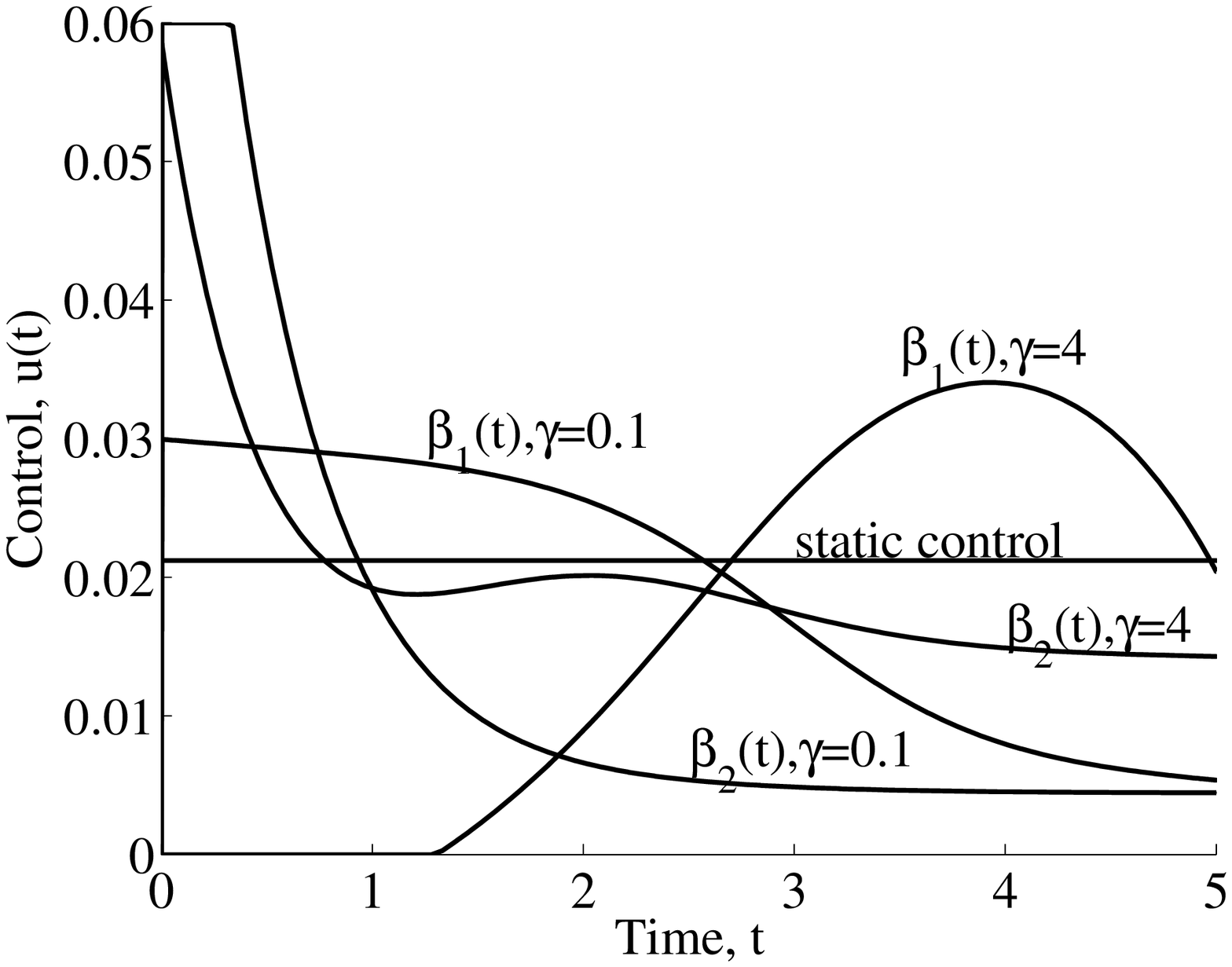} }
\hfill
\subfloat[Cumulative resource spent by the optimal and static controls (variable spreading rate). \label{fig:cumulative_resource_spent_variable_beta}]{
\includegraphics[width=80mm]{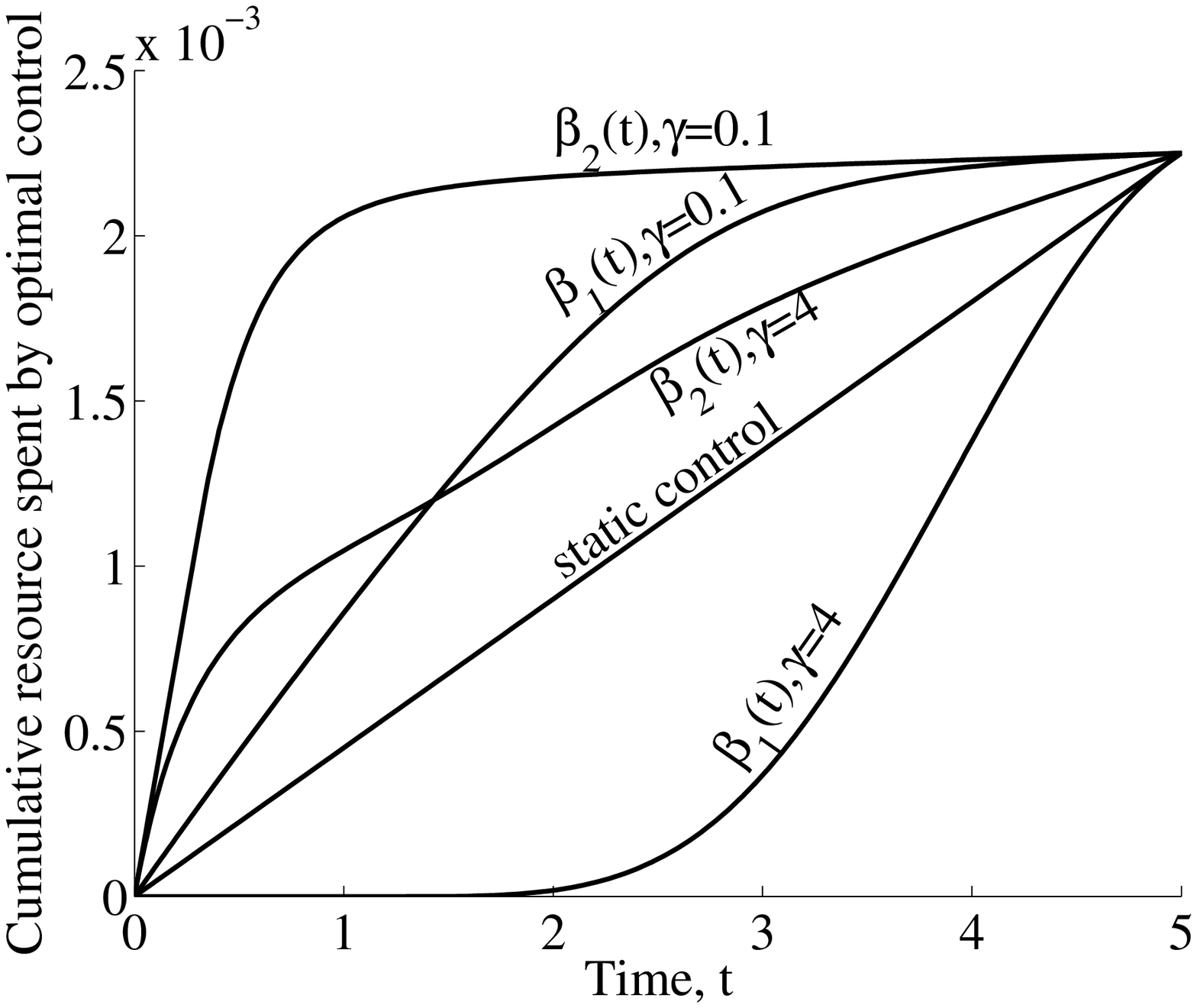} }
\caption{Parameter values: $c(u(t))=u^2(t),T=5,u_{max}=0.06,B=u_{max}^2T/8,s_0=0.01, \alpha = 0.5$.}
\label{fig:control_cumulative_resource_variable_beta}
\end{figure}

Often in practical scenarios, the interest level of the population in talking about the subject of the campaign varies with time. In the case of a political campaign, interest of people in talking about polls will increase as the election date approaches. Similarly, as a product or a movie becomes old, people lose interest in them. We have modeled these two scenarios by using increasing and decreasing spreading rate profiles given by the following equations:
\begin{subequations}
\label{eq:var_beta}
\begin{eqnarray}
\beta_1(t) & = & \beta_m + \left( \frac{\beta_M-\beta_m}{1+e^{-a_1(t-c_1)}} \right), \label{eq:var_beta1} \\
\beta_2(t) & = & (\beta_M - \beta_m) \left(1 - \frac{1}{1+e^{-a_2(t-c_2)}} \right). \label{eq:var_beta2}
\end{eqnarray}
\end{subequations}
The parameter values are set to: $\beta_m=0.01, \beta_M=2, T=5, a_1=2, c_1=3, a_2=2, c_2=2$ and $t\in[0,5]$. The spreading rate profiles are shown in Fig. \ref{fig:variable_beta}.

The shapes of the control signals and cumulative resource expenditure, for the spreading rate profiles given in Fig. \ref{fig:variable_beta}, are shown in Fig. \ref{fig:control_cumulative_resource_variable_beta} for two different values of recovery rate, $\gamma$=0.1 and 4. We can see that the spreading rate profile affects the optimal control. When the recovery rate is set to a small value $\gamma=0.1$, for the decreasing profile $\beta_2(t)$, the optimal control is strong in the beginning stages and gradually tapers off. But for the increasing profile $\beta_1(t)$, the control is relatively smaller in the beginning stages. The reason for such a behavior is the small value of spreading rate, in the case of increasing profile $\beta_1(t)$, in the beginning stages, which will not aid further information dissemination in that period of slow spreading. At later stages spreading is stronger, so some resource is saved to be utilized in middle and final stages by the optimum strategy. When the recovery rate is raised to $\gamma=4$, we see a behavior which is, in principle, same as in the case of fast recovery in Sec. \ref{sec:result_const_beta}. The control effort is shifted to middle and final stages of the campaign in the cases of both $\beta_1(t)$ and $\beta_2(t)$ for $\gamma=4$ when compared to the case of $\gamma=0.1$. This facilitates in having a sizeable population of spreaders throughout the campaign horizon which keeps dwindling due to fast recovery.

\subsection{Effect of various parameters}
\label{sec:result_effect_param}


\begin{figure}[ht!]
\subfloat[Cost, $J$ vs. budget, $B$. \label{fig:cost_vs_budget}]{
\includegraphics[width=80mm]{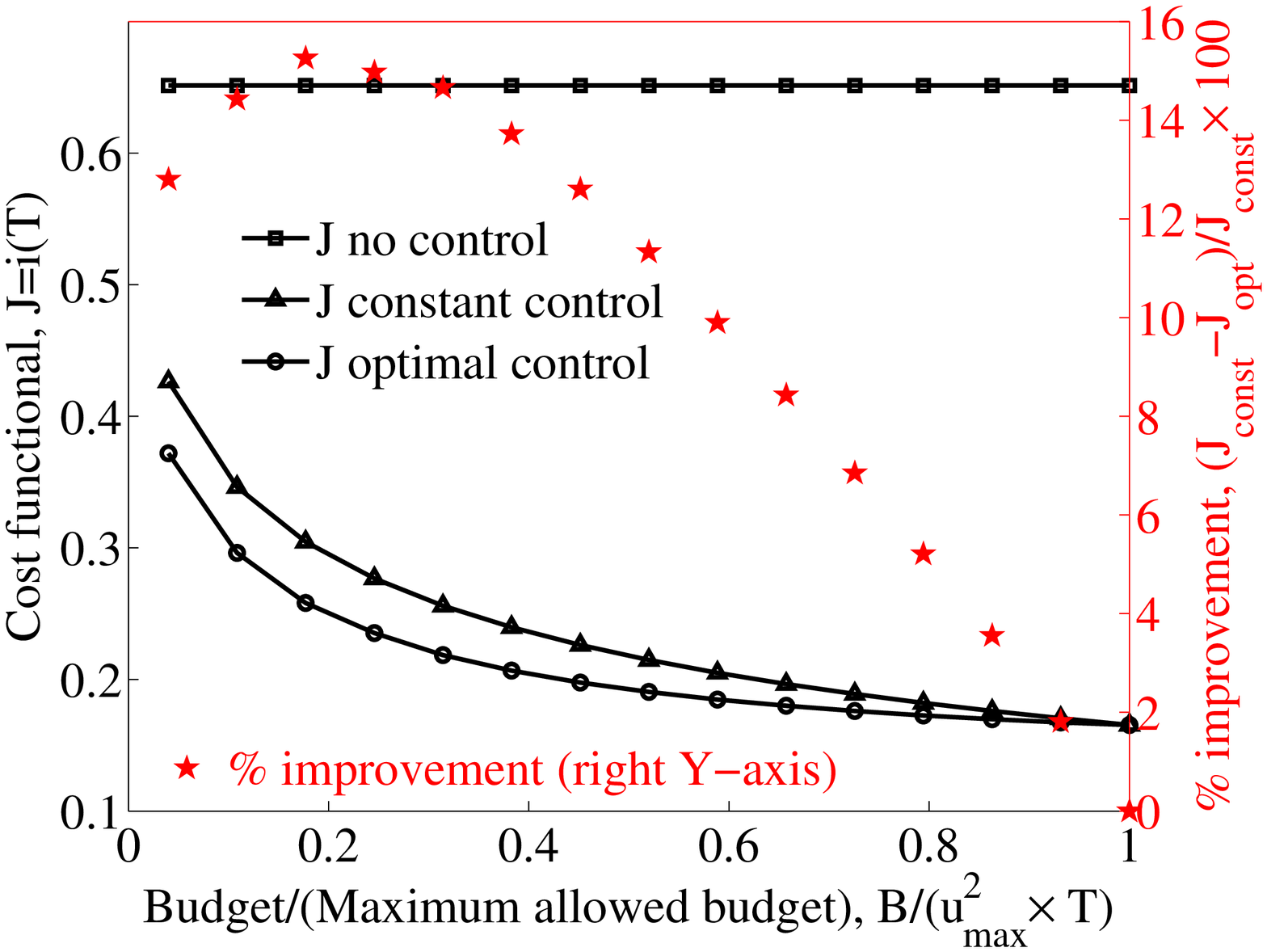} }
\hfill
\subfloat[Shapes of optimal and static controls for different budget values. \label{fig:control_opt_stat_with_B}]{
\includegraphics[width=78mm]{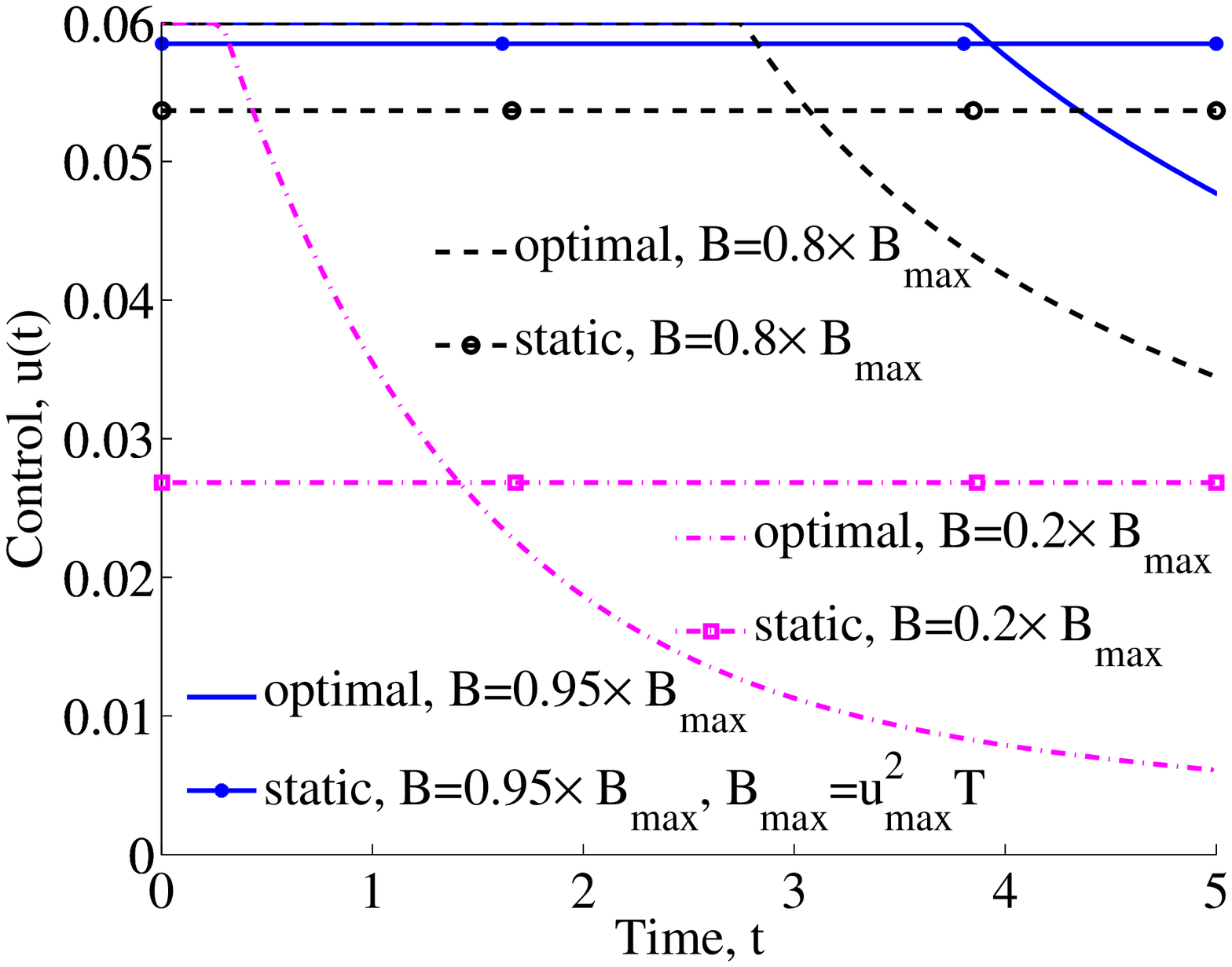} }
\caption{Parameter values: $c(u(t))=u^2(t), \beta=0.8, \gamma=0.1, T=5, u_{max}=0.06, s_0=0.01, \alpha = 0.5$.}
\label{fig:cost_vs_budget_and_shapes_controls_wrt_B}
\end{figure}

\begin{figure}[ht!]
\centering
\includegraphics[width=85mm]{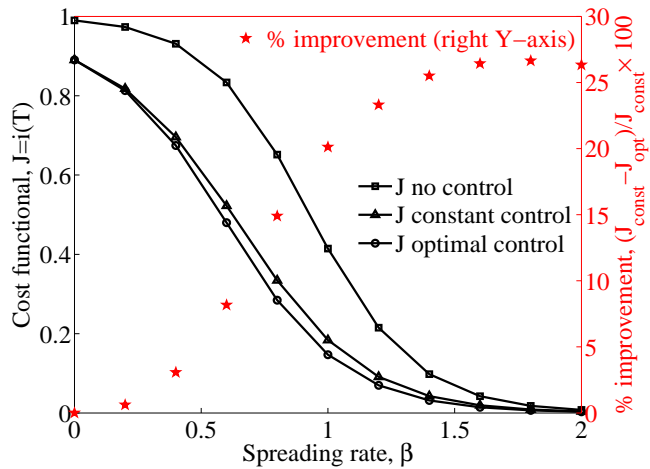}
\caption{Cost, $J$ vs. spreading rate, $\beta$. Parameter values: $c(u(t))=u^2(t), \gamma=0.1, T=5, u_{max}=0.06, B=u_{max}^2T/8, s_0=0.01, \alpha = 0.5$.}
\label{fig:cost_vs_beta}
\end{figure}

\begin{figure}[ht!]
\centering
\includegraphics[width=85mm]{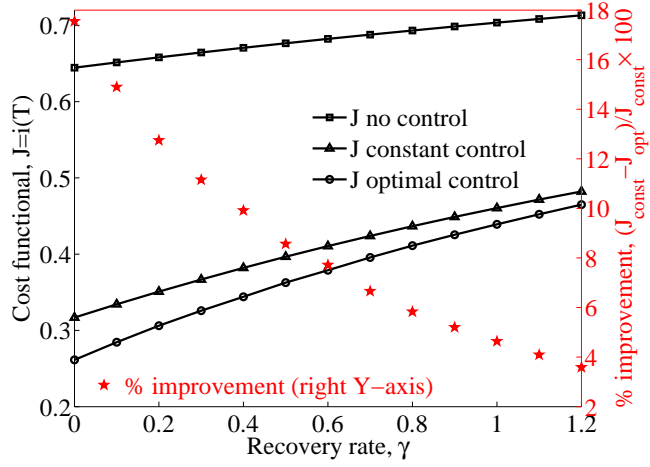}
\caption{Cost, $J$ vs. recovery rate, $\gamma$ Parameter values: $c(u(t))=u^2(t), \beta=0.8, T=5, u_{max}=0.06, B=u_{max}^2T/8, s_0=0.01, \alpha = 0.5$.}
\label{fig:cost_vs_gamma}
\end{figure}

\begin{figure}[ht!]
\centering
\includegraphics[width=85mm]{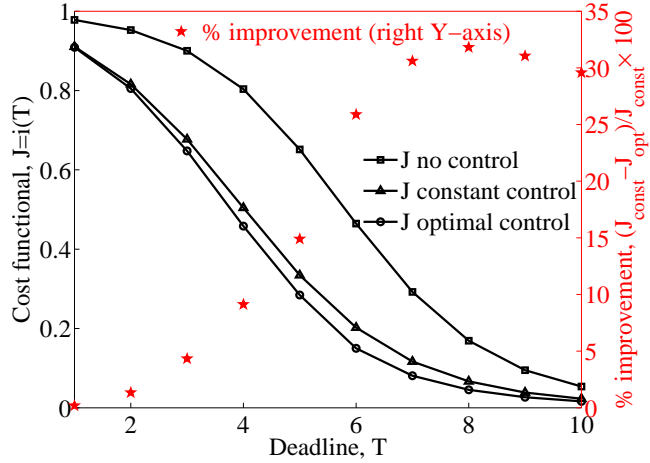}
\caption{Cost, $J$ vs. campaign deadline, T. Parameter values: $c(u(t))=u^2(t), \beta=0.8, \gamma=0.1, u_{max}=0.06, B=u_{max}^2(5)/8, s_0=0.01, \alpha = 0.5$.}
\label{fig:cost_vs_T}
\end{figure}

\begin{figure}[ht!]
\centering
\includegraphics[width=85mm]{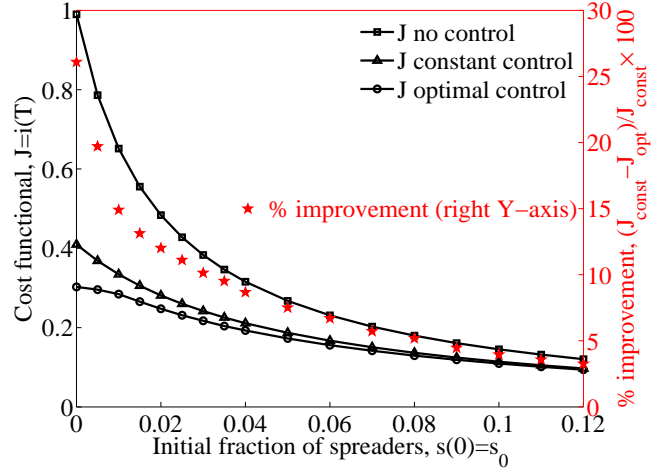}
\caption{Cost, $J$ vs. initial number of spreaders, $s_0$. Parameter values: $c(u(t))=u^2(t), \beta=0.8, \gamma=0.1, T=5, u_{max}=0.06, B=u_{max}^2T/8, \alpha = 0.5$.}
\label{fig:cost_vs_s0}
\end{figure}

\begin{figure}[ht!]
\centering
\includegraphics[width=85mm]{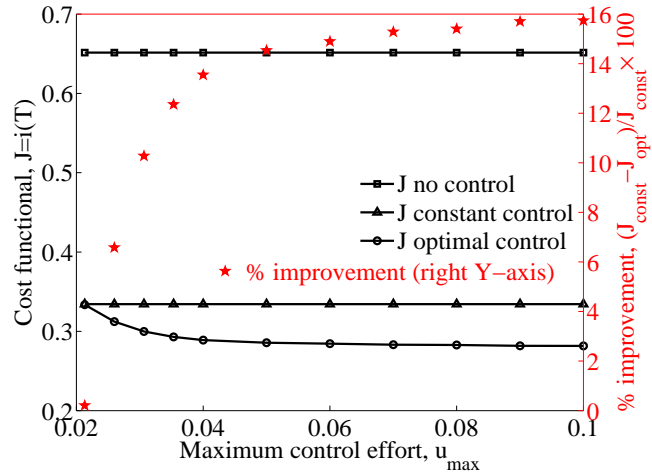}
\caption{Cost, $J$ vs. maximum control effort, $u_{max}$. Parameter values: $c(u(t))=u^2(t), \beta=0.8, \gamma=0.1, T=5, B=(0.06)^2\times T/8, s_0=0.01, \alpha = 0.5$.}
\label{fig:cost_vs_umax}
\end{figure}

\begin{figure}[ht!]
\centering
\includegraphics[width=85mm]{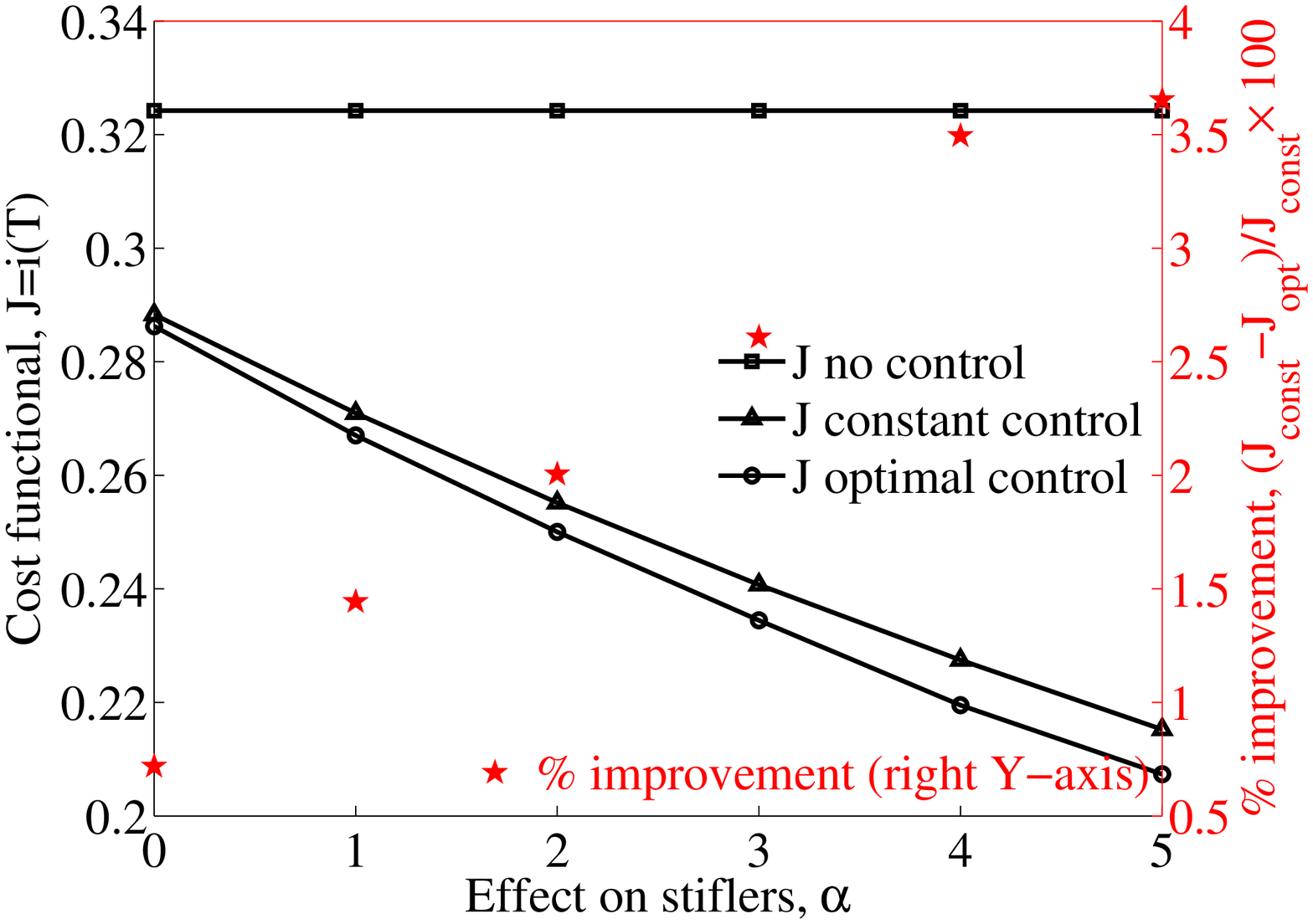}
\caption{Cost, $J$ vs. effectiveness on stiflers, $\alpha$. Parameter values: $c(u(t))=u^2(t), \beta=4, \gamma=6, T=5, u_{max}=0.06, B=u_{max}^2T/8, s_0=0.01$.}
\label{fig:cost_vs_alpha}
\end{figure}

In this section we consider a spreading rate which is constant over the whole campaign duration, \emph{i.e.} $\beta(t)=\beta$, for $t\in[0~~T]$, and study the effect of various model parameters on the cost function defined in Eq. (\ref{eq:cost_funtional}). We have also compared the cost achieved by the optimal control to that achieved by the static control and show significant performance gains for certain parameter values.

From Fig. \ref{fig:cost_vs_budget}, the performance gains due to optimal control compared to static control are significant only for low and intermediate budget values. The shape of the optimal control increasingly resembles the static control for high budget values (Fig. \ref{fig:control_opt_stat_with_B}). In both the cases, for the maximum value of the budget ($u_{max}^2T$), the strategy is to apply maximum control strength $u_{max}$ throughout $[0~~T]$, so the performance is also the same. For values of the budget close to the maximum possible, both strategies lead to controls of similar shape and hence have similar performances.

From Fig. \ref{fig:cost_vs_beta}, as the epidemic becomes more viral, the relative improvement offered by the optimal control over the static control increases. However, at extremely high spreading rate, for a given recovery rate, the benefit offered by campaigning (optimal or static) compared to no campaign scenario is negligible. For the political campaigning example discussed in Sec. \ref{sec:system_model_prob_formulation}, this means that we gain in the fraction of spreaders/stiflers as people are more and more interested in talking about upcoming elections. However, if the interest becomes very high, the absolute difference in the fraction of spreaders/stiflers is almost negligible compared to the case when no campaigning is used. This is so because the spreading rate is so high that we are able to reach almost the whole population even without campaigning.

From Fig. \ref{fig:cost_vs_gamma}, increasing the recovery rate for a fixed spreading rate has qualitatively the same effect as decreasing spreading rate for a fixed recovery rate. Once someone recovers and becomes a stifler, she does not disseminate the information (unless recruited by the control). With a higher recovery rate, a larger fraction of spreaders becomes stiflers quickly. They lose interest and do not spread the information actively. Due to diminished information dissemination, the fraction of ignorants on the election day increases. This is true for any control strategy, even the `no control' strategy, as the three plots in Fig. \ref{fig:cost_vs_gamma} show. It is observed also that the relative performance improvement achieved by the optimal strategy over the static strategy reduces.

From Fig. \ref{fig:cost_vs_T}, for given $\beta$ and $\gamma$, as the deadline increases, the relative performance improvement of optimal control compared to static control increases. But for large deadlines, it is possible to reach more people compared to the case of shorter deadline, for chosen values of $\beta$ and $\gamma$. This fraction is close to the equilibrium value which the chosen values of $\beta$ and $\gamma$ allow. Hence, the advantage offered by campaigning (optimal or static) over no campaign strategy decreases. If the deadline is larger, the fraction of people who remain ignorant on the polling day is expected to reduce, simply because there is more time for the information to spread (to its equilibrium value).

Fig. \ref{fig:cost_vs_s0} reveals that optimal campaigning offers benefit over static campaigning only if we start the campaign early, \emph{i.e.}, when number of spreaders in the population at $t=0$, represented by $s_0$, is low. This means that one should not bother about calculating/implementing an optimal campaigning strategy if a sizeable population already knows the message at the beginning of the campaign---the static control strategy is good enough. For practical scenarios, low $s_0$ is usually satisfied, hence the optimal strategy is expected to outperform the static strategy by a considerable margin. More often than not, we know about the information as a result of the ongoing campaign.

We plot the variation of $J$ with respect to maximum control effort $u_{max}$ in Fig. \ref{fig:cost_vs_umax}. As expected, the improvement of the optimal control over the static control gradually increases and then saturates. The minimum allowed value of $u_{max}$ is the value of the static control. As $u_{max}$ increases the system benefits from stronger early control effort, which is beneficial for parameter values chosen. But for too high values of $u_{max}$ the optimal control never saturates to the maximum value and hence further improvement in the cost is not possible by increasing $u_{max}$. Fig. \ref{fig:cost_vs_umax} studies the effect of $u_{max}$ on $J$, hence other parameters, specifically $B$ and $T$ are fixed. Thus the value of the static control, $u_{stat}=\sqrt{B/T}$ is fixed throughout the figure.

Fig. \ref{fig:cost_vs_alpha} shows variation of the cost function with respect to $\alpha$. As expected, increasing $\alpha$ leads to reduction in fraction of ignorants on the polling day. It increases the conversion of stiflers to spreaders, which enhances information dissemination. Higher values of $\alpha$ can be attained by better designed advertisements. In addition, we observe that increasing $\alpha$ leads to increase in relative improvement of the cost function achieved by the optimal strategy compared to the static campaigning strategy.

\section{Conclusion}
\label{sec:conclusion}
In this work we have formulated an optimal control problem to maximize the spread of information under a fixed campaigning budget constraint. The information spread dynamics is assumed to follow the Maki Thompson rumor model, which is more suitable in this context than SIS/SIR epidemic models used in some of the previous studies. The control signal converts ignorants and stiflers into spreaders. This can be done via strategies such as advertising in mass media, publishing manifestos, door-to-door campaigns etc., depending upon the application---election, product promotion, crowdfunding, social awareness campaigns, to mention a few. We assume general non-linear campaigning costs and show the existence of a solution to the formulated optimal control problem. Note that the standard Filippov/Cesari theorems are not applicable in this situation. We solve the optimal control problem using Pontryagin's Minimum Principle and a modified version of forward backward sweep technique for numerical computation, to accommodate the isoperimetric budget constraint in our formulation. The techniques developed in this paper are general and can be applied to other similar optimal control problems.

To model practical situations, such as increasing interest of people in talking about elections as polling day approaches or diminishing interest in a movie after its release, we have allowed the spreading rate profile of the information epidemic to vary during the campaign duration. We have studied the shape of the optimal control signal for different model parameters and spreading rate profiles. Variations of the optimal campaigning costs with respect to various model parameters are also studied and results compared with the static campaigning strategy. In the static strategy the control is constant throughout the decision horizon and respects the same budget constraint as the optimal strategy. We have found that the optimal strategy achieves significant performance improvements compared to the static strategy for a wide range of model parameters.




\bibliographystyle{model3-num-names}
\bibliography{bibliography_database}


\end{document}